\documentclass[fdp,a4paper,fleqn%
]{w-art}
\usepackage{times,cite,w-thm}
\theoremstyle{plain}

\theoremstyle{definition}

\usepackage[]{graphicx}

\begin{document}

\title{Formation \& evolution of cosmic superstrings: a short review}

\author{M. Sakellariadou
  \footnote{E-mail:~\textsf{Mairi.Sakellariadou@kcl.ac.uk}, Phone:
    +44\,(0)20\,7848\,1535, 
Fax: +44\,(0)20\,7848\,2420}}
\address
{Physics Department, King's College London, University of London,
  Strand WC2R 2LS, United Kingdom}

\begin{abstract}
I will briefly review the formation and evolution of cosmic
superstrings, in the context of brane-world cosmological models within
M-theory. These objects can play the r\^ole of cosmic strings,
offering a variety of astrophysical consequences, which I will
briefly discuss.
\end{abstract}

\maketitle                   

Current cosmological and astrophysical data, and in particular,
measurements of the Cosmic Microwave Background (CMB) temperature
anisotropies, strongly support the inflationary paradigm.  However,
despite its success, inflation remains a paradigm in search of a
model. As the cosmological data keep improving impressively fast, it
becomes urgent to find an inflationary model with a solid theoretical
foundation. Since
studies~\cite{Calzetta:1992gv,Calzetta:1992bp,Germani:2007rt} on the
probability of the onset of inflation indicate that it should take
place in the deep quantum gravity regime, it is natural to study
inflation in the process of brane interactions, within brane cosmology
in string theory.  If M-theory is indeed the theory of
everything, it should provide a natural inflationary scenario.  In
this approach, one will identify the inflaton and its properties,
while cosmological measurements will help to determine the precise
stringy description of our universe.  There is a number of
cosmological models motivated by string theory.  Compactification to
four space-time dimensions leads to scalar fields and moduli, which
could play the r\^ole of the inflaton field, provided they do not roll
quickly; one has to provide a mechanism for moduli stabilization.
Brane annihilations allow the survival only of three-dimensional
branes~\cite{Durrer:2005nz,Nelson:2008sv} with the production of
fundamental string (F-strings) and one-dimensional Dirichlet branes
(D-strings). Thus, brane inflation~\cite{Jones:2002cv} leads to the
formation of cosmic superstrings, in an analogous way to cosmic
strings, which are generically formed~\cite{Jeannerot:2003qv} at the
end of hybrid inflation in the context of supersymmetric grand
unified theories.  In what follows, I will shortly review the
formation and evolution of cosmic superstrings and briefly discuss
their astrophysical
consequences~\cite{Davis:2005dd,Sakellariadou:2008ie,Sakellariadou:2009ev,Copeland:2009ga}.

To illustrate the
formation~\cite{Sarangi:2002yt,Jones:2003da,Dvali:2003zj} of cosmic
superstrings at the end of brane inflation, consider a D$p$-D$\bar{p}$
brane-anti-brane pair annihilation to form a D$(p-2)$ brane.  Each
brane has a U(1) gauge symmetry and the gauge group of the pair is
U(1)$\times$U(1). The daughter brane possesses a U(1) gauge group,
which is a linear combination of the two original U(1)'s. The branes
move towards each other and as their inter-brane separation decreases
below a critical value, the tachyon field, which is an open string
mode stretched between the two branes, develops an instability.  The
rolling of the tachyon field leads to the decay of the parent
branes. Tachyon rolling leads to spontaneously symmetry breaking,
which supports defects with even co-dimension. So, brane annihilation
leads to vortices, D-strings, produced via the Kibble mechanism. The
other linear combination disappears, since only one brane remains
after the brane collision; it is thought to disappear by having its
fluxes confined by fundamental closed strings. Cosmic superstrings are
of cosmological size and they could play the r\^ole of cosmic strings.

It is worth mentioning that cosmic superstrings are not an inevitable
consequence of brane inflation. There are models in which these
objects do not appear, {\sl e.g.}, inflationary models based on the
condensation of an open string tachyon, or based on closed string
moduli. However, there are also scenarios leading to cosmic
superstring formation that do not rely on brane inflation, {\sl e.g.},
the Hagedorn phase transition. The density of string states increases
exponentially with their mass~\cite{hag1} and at a particular scale
(the Hagedorn scale, close to the string scale), there is a phase
transition and all the available energy goes into creating long
(super-horizon) strings, as opposed to (sub-horizon) string loops. It
is reasonable to expect that some of these objects survive until
now. Thus, if inflation is not required, one can
consider~\cite{hag2,Biswas:2008ti} starting with a large universe in
the Hagedorn phase, within the string gas
scenario~\cite{sg1,sg2,Battefeld:2005av}.

There is a number of differences between solitonic cosmic strings and
cosmic superstrings.  Cosmic strings are classical objects, assumed to
share the characteristics of type-II vortices in the Abelian Higgs
model.  Cosmic superstrings, despite the fact that they are
cosmologically extended, they are quantum objects.  Numerical
simulations~\cite{ns1,ns2} of type-II strings in the Abelian Higgs
model suggest that the probability that a pair of strings will
reconnect, after they intersect, is close to unity.  The reconnection
probability for cosmic superstrings is smaller (often much smaller)
than unity. The corresponding intercommutation probabilities,
calculated in string perturbation theory, depend on the type of
strings and on the details of compactification. For fundamental
strings, reconnection is a quantum process and takes place with a
probability of order $g_{\rm s}^2$ (where $g_{\rm s}$ denotes the
string tension).  It can thus be much less than one, leading to an
increased density of strings~\cite{Sakellariadou:2004wq}, which
implies an enhancement of various observational signatures.  Even
though the value of $g_{\rm s}$ and the scale of the confining
potential, which determine the reconnection probability, are not
known, it was found~\cite{Jackson:2004zg} that for a large number of
models the reconnection probability for F-F collisions lies in the
range between $10^{-3}$ and 1; for D-D collisions is anything between
0.1 to 1; for F-D collisions it can vary from 0 to 1.  

Brane collisions can also
produce~\cite{Copeland:2003bj,Leblond:2004uc} bound states,
$(p,q)$-strings, which are composites of $p$ F-strings and $q$
D-strings, with $p, q$ relatively prime integers.  The presence of
stable bound states implies the existence of $Y$-junctions, where two
different types of string meet at a point and form a bound state
leading away from that point.

The tension of solitonic strings is set from the energy scale of the
phase transition followed by a spontaneously broken symmetry which
left behind strings as false vacuum remnants.  Cosmic superstrings
span a whole range of tensions, set from the particular brane
inflation model employed.  The tension of F-strings in 10 dimensions
is $\mu_{\rm F}=1/(2\pi\alpha')$, and the tension of D-strings is
$\mu_{\rm D}=1/(2\pi\alpha' g_{\rm s})$.  In 10 flat dimensions,
supersymmetry dictates that the tension of the $(p,q)$ bound states
reads~\cite{tension}
\begin{equation}
\mu_{(p,q)}=\mu_{\rm F}\sqrt{p^2 + q^2/g_{\rm s}^2}~.
\label{squarelaw-tension}
\end{equation}
Individually, F- and D-strings are ${1\over 2}$-BPS
(Bogomol'nyi-Prasad-Sommerfield) objects, each breaking a different
half of the supersymmetry. Equation (\ref{squarelaw-tension})
represents the BPS bound for an object carrying the charges of $p$ F-
and $q$ D-strings.  In IIB string theory, where our universe can be
described as a brane-world scenario with flux compactification, the
string tension is different ({\it see e.g.}, Ref.~\cite{Firouzjahi:2006vp})
from the (simple) expression given in Eq.~(\ref{squarelaw-tension})
and depends on the particular choice of flux compactification.

The evolution of cosmic superstrings is a very complex issue, which
depends on the brane inflation model. Let me first briefly summarise
our understanding of the evolution of cosmic string
networks~\cite{Sakellariadou:2006qs,Sakellariadou:2007bv}.

The first (analytical) studies of the evolution of cosmic string
networks indicated~\cite{scaling} the existence of {\sl scaling}, in
the sense that at least the basic properties of the network can be
characterised by a single length scale, roughly the persistence length
or the inter-string distance $\xi$ which grows with the horizon.  The
scaling solution was supported~\cite{scal1} by subsequent numerical
work; further investigation revealed~\cite{scal2,scal3} however the
existence of dynamical processes at scales much smaller than $\xi$.

If the super-horizon strings are characterised by a single length
scale $\xi(t)$, the typical distance between the nearest string
segments and the typical curvature radius of the strings are of the
order of
\begin{equation}
\xi(t)=\Big({\rho_{\rm super-horizon}\over\mu}\Big)^{-1/2}=\kappa^{-1/2} t~.
\end{equation}
Early numerical simulations confirmed that the typical curvature
radius of long strings and the characteristic distance between them
are both comparable to the evolution time, while they found the
existence of an important small-scale superimposed on the
super-horizon strings~\cite{scal3}.

The sub-horizon loops, their size distribution, and the mechanism of
their formation remained for years the least understood part of the
string evolution. Recent numerical simulations,
found~\cite{Ringeval:2005kr} evidence of a scaling regime for the
cosmic sub-horizon string loops in the radiation- and matter-dominated
eras down to the hundredth of the horizon time, a result which has
been confirmed~\cite{Polchinski:2006ee} by analytical studies.

The evolution of cosmic superstrings is a much more involved issue.
Cosmic superstring networks have not only sub-horizon and
super-horizon strings, but also $Y$-junctions, which {\it a priori}
may prevent a scaling solution. Moreover, one must consider a
multi-tension spectrum and reconnection probabilities which can be
much lower that unity. Certainly, computers are at present much more
efficient than in the eighties and nineties when we performed the
first numerical experiments with solitonic strings, and we obviously
gained a lot of experience from those studies. Nevertheless, one must
not forget that evolution of cosmic strings has been almost
exclusively studied in the (simple but unrealistic) case of the infinitely
thin approximation.

A number of numerical
experiments~\cite{Sakellariadou:2004wq,sim1,sim2,sim3,sim4,sim5,Rajantie:2007hp,Urrestilla:2007yw,Sakellariadou:2008ay,Bevis:2009az,Avgoustidis:2009ke}
have attempted to get some insight into the evolution of cosmic
superstring networks, and in particular, investigate whether the bound
$(p,q)$ states may obstruct the existence of a scaling solution.  In
Ref.~\cite{Sakellariadou:2008ay} for instance, the evolution of a
cosmic string network was studied via numerical experiments in a
simple field theory model of bound states (in an analogy to the
Abelian Higgs model) which however incorporates the main features of
string theory. It was found~\cite{Sakellariadou:2008ay} robust
evidence for scaling of all three components --- $p$ F-strings, $q$
D-strings, $(p,q)$ bound states --- independently of initial
conditions. Scaling was confirmed from all other numerical studies.

At first, cosmic strings were regarded as completely different objects
than cosmic superstrings.  Cosmic strings on the one hand, stretch
across cosmological distances and though exceedingly thin, they are
sufficiently massive to have important gravitational effects. Cosmic
superstrings on the other hand, were considered as being far too small
to have any directly observable effects. This view has however
considerably changed over the last few years, and it is believed that
under certain circumstances, cosmic superstrings can, in the context
of brane-world cosmological models, grow to macroscopic sizes and play
the r\^ole of cosmic strings.  Cosmic strings and cosmic superstrings
can produce a variety of astrophysical signatures, including
gravitational waves~\cite{gw1,gw2,gw3}, ultra high energy cosmic
rays~\cite{uhecr}, gamma ray bursts~\cite{grb}, radio
bursts~\cite{rb}, cosmic 21 cm power spectrum~\cite{cps},
magnetogenesis~\cite{magneto}, CMB at small angular
scales~\cite{cmb1}, CMB polarization, and
lensing~\cite{lensing1,lensing2}. Their signatures though are quite
distinct, due to the differences of the two networks, discussed
earlier. In particular, since non-periodic F-strings ending on
D-branes are accompanied~\cite{Davis:2008kg} by the formation of {\sl
  cusps}, it can be demonstrated~\cite{Davis:2008kg} that pairs of
Y-junctions, such as would form after intercommutations of F- and
D-strings, generically contain cusps. This feature of cosmic
superstrings opens up the possibility of extra channels of energy loss
from a cosmic superstring network.

Unfortunatley, up to now cosmic strings as well as cosmic superstrings
remain in the realm of theory, as hypothetical objects awaiting for an
observational verification. Observational support of cosmic strings
will confirm the validity of phase transtitions in the context of
grand unified theories applied in the early universe
cosmology. Observational support of cosmic superstrings will justify
the validity of string theory and will shed some light in the
appropriate stringy description of our universe. Ongoing theoretical
research will eventually unravel the evolution of cosmic superstring
networks, while astrophysical observations will provide the means
which may falsify the theory or enlighten the theoretical models.

\begin{acknowledgement}
It is a pleasure to thank the organisers of the workshop on
Cosmology-Strings: Theory-Cosmo\-lo\-gy-Phenomenology, {\sl European
  Institute for scienes and their applications}
(http://www.physics.ntua.gr/corfu2009/), for inviting me to
give this talk, in the beautiful island of Corfu in Greece. This work
is partially supported by the European Union through the Marie Curie
Research and Training Network UniverseNet (MRTN- CT-2006-035863).
\end{acknowledgement}


\begin{thebibliography}{[1]}

\bibitem{Calzetta:1992gv}
  E.~Calzetta and M.~Sakellariadou,
  Phys.\ Rev.\  D {\bf 45}, 2802 (1992).

\bibitem{Calzetta:1992bp}
  E.~Calzetta and M.~Sakellariadou,
  Phys.\ Rev.\  D {\bf 47}, 3184 (1993).

\bibitem{Germani:2007rt}
  C.~Germani, W.~Nelson and M.~Sakellariadou,
  Phys.\ Rev.\  D {\bf 76}, 043529 (2007).

\bibitem{Durrer:2005nz}
  R.~Durrer, M.~Kunz and M.~Sakellariadou,
  Phys.\ Lett.\  B {\bf 614}, 125 (2005).

\bibitem{Nelson:2008sv}
  W.~Nelson and M.~Sakellariadou,
  Phys.\ Lett.\  B {\bf 674}, 210 (2009).

\bibitem{Jones:2002cv}
  N.~T.~Jones, H.~Stoica and S.~H.~H.~Tye,
  JHEP {\bf 0207}, 051 (2002).

\bibitem{Jeannerot:2003qv}
  R.~Jeannerot, J.~Rocher and M.~Sakellariadou,
  Phys.\ Rev.\  D {\bf 68}, 103514 (2003).

\bibitem{Davis:2005dd}
  A.~C.~Davis and T.~W.~B.~Kibble,
  Contemp.\ Phys.\  {\bf 46}, 313 (2005).

\bibitem{Sakellariadou:2008ie}
  M.~Sakellariadou,
  Phil.\ Trans.\ Roy.\ Soc.\ Lond.\  A {\bf 366}, 2881 (2008).

\bibitem{Sakellariadou:2009ev}
  M.~Sakellariadou,
  Nucl.\ Phys.\ Proc.\ Suppl.\  {\bf 192-193}, 68 (2009).

\bibitem{Copeland:2009ga}
  E.~J.~Copeland and T.~W.~B.~Kibble,
  arXiv:0911.1345 [hep-th].

\bibitem{Sarangi:2002yt}
  S.~Sarangi and S.~H.~H.~Tye,
  Phys.\ Lett.\  B {\bf 536}, 185 (2002).

\bibitem{Jones:2003da}
  N.~T.~Jones, H.~Stoica and S.~H.~H.~Tye,
  Phys.\ Lett.\  B {\bf 563}, 6 (2003).

\bibitem{Dvali:2003zj}
  G.~Dvali and A.~Vilenkin,
  JCAP {\bf 0403}, 010 (2004).

\bibitem{hag1}
M.~Sakellariadou and A.~Vilenkin, Phys.\ Rev.\ D {\bf 37}, 885 (1988).

\bibitem{hag2}
A.~Nayeri, R.~H.~Brandenberger and C.~Vafa, Phys.\ Rev.\ Lett. {\bf 97}, 
021302 (2006).

\bibitem{Biswas:2008ti}
  T.~Biswas,
  arXiv:0801.1315 [hep-th].

\bibitem{sg1}
R.~H.~Brandenberger and C.~Vafa, Nucl.\ Phys.\ B {\bf 316}, 391 (1989).

\bibitem{sg2}
M.~Sakellariadou, Nucl.\ Phys.\ B {\bf 468}, 319 (1996).

\bibitem{Battefeld:2005av}
  T.~Battefeld and S.~Watson,
  Rev.\ Mod.\ Phys.\  {\bf 78}, 435 (2006).

\bibitem{ns1}
E.~P.~S.~Shellard, Nucl.\ Phys. B\,
{\bf 283}, 624 (1987).

\bibitem{ns2}
P.~Laguna and R.~A.~Matzner, Phys.\ Rev. D\,
{\bf 41}, 1751 (1990).

\bibitem{Sakellariadou:2004wq}
  M.~Sakellariadou,
  JCAP {\bf 0504}, 003 (2005).

\bibitem{Jackson:2004zg}
  M.~G.~Jackson, N.~T.~Jones and J.~Polchinski,
  JHEP {\bf 0510}, 013 (2005).

\bibitem{Copeland:2003bj}
  E.~J.~Copeland, R.~C.~Myers and J.~Polchinski,
  JHEP {\bf 0406}, 013 (2004).

\bibitem{Leblond:2004uc}
  L.~Leblond and S.~H.~H.~Tye,
  JHEP {\bf 0403}, 055 (2004).

\bibitem{tension} J.~H.~Schwarz, Phys.\ Lett.\ B\,
  {\bf 360}, 13 (1995).

\bibitem{Firouzjahi:2006vp}
  H.~Firouzjahi, L.~Leblond and S.~H.~H.~Tye,
JHEP  {\bf 0605}, 047 (2006).

\bibitem{Sakellariadou:2006qs}
  M.~Sakellariadou,
  Lect.\ Notes Phys.\  {\bf 718}, 247 (2007).

\bibitem{Sakellariadou:2007bv}
  M.~Sakellariadou,
  Lect.\ Notes Phys.\  {\bf 738}, 359 (2008).

\bibitem{scaling}
T.~W.~B.~Kibble, Nucl.\ Phys. B, {\bf 252}, 227 (1985); Erratum-ibid.,
{\bf 261}, 750 (1985).

\bibitem{scal1}
A.~Albrecht and N.~Turok, Phys.\ Rev.\ Lett., {\bf 54}, 1868 (1985).

\bibitem{scal2}
D.~P.~Bennett and F.~R.~Bouchet, Phys.\ Rev.\ Lett., {\bf 60}, 257 (1988).

\bibitem{scal3}
M.~Sakellariadou and A.~Vilenkin, Phys.\ Rev.\ D, {\bf 42}, 349 (1990).

\bibitem{Ringeval:2005kr} C.~Ringeval, M.~Sakellariadou and
  F.~Bouchet,
  JCAP {\bf 0702}, 023 (2007).

\bibitem{Polchinski:2006ee}
  J.~Polchinski and J.~V.~Rocha,
  Phys.\ Rev.\  D {\bf 74}, 083504 (2006).

\bibitem{sim1} A.~Avgoustidis and E.~P.~S. Shellard, Phys.\ Rev.\ D,
  {\bf 71}, 123513 (2005).

\bibitem{sim2} E.~Copeland and P.~M.~Saffin,
JHEP, {\bf 0511}, 023 (2005).

\bibitem{sim3}
P.~M.~Saffin,  2005, JHEP, {\bf 0509}, 011 (2005).

\bibitem{sim4} A.~Avgoustidis and E.~P.~S. Shellard, Phys.\ Rev.\ D,
  {\bf 73}, 041301 (2006).

\bibitem{sim5}
M.~Hindmarsh and P.~M.~Saffin, JHEP, {\bf 0608}, 066 (2006).

\bibitem{Rajantie:2007hp}
  A.~Rajantie, M.~Sakellariadou and H.~Stoica,
  JCAP {\bf 0711}, 021 (2007).

\bibitem{Urrestilla:2007yw}
  J.~Urrestilla and A.~Vilenkin,
  JHEP {\bf 0802}, 037 (2008).

\bibitem{Sakellariadou:2008ay}
  M.~Sakellariadou and H.~Stoica,
  JCAP {\bf 0808}, 038 (2008).

\bibitem{Bevis:2009az}
  N.~Bevis, {\sl et. al.}, 
  Phys.\ Rev.\  D {\bf 80}, 125030 (2009).

\bibitem{Avgoustidis:2009ke}
  A.~Avgoustidis and E.~J.~Copeland,
  arXiv:0912.4004 [hep-ph].

\bibitem{gw1} T.~ Damour and A.~Vilenkin, Phys.\ Rev.\ D {\bf 71},
  063510 (2005).

\bibitem{gw2} 
J.~Hogan, Phys.\ Rev.\ D {\bf 74}, 043526 (2006).

\bibitem{gw3} X.~Siemens, V.~Mandic and J.~Creighton,
  Phys.\ Rev.\ Lett.\ {\bf 98}, 111101 (2007).

\bibitem{uhecr} V.~Berezinsky, B.~Hnatyk and A.~Vilenkin,
  Phys.\ Rev.\ D {\bf 64}, 043004 (2001).

\bibitem{grb} P.~Bhattacharjee and G.~Sigl, Phys.\ Rept.\ {\bf 327},
  109 (2000).

\bibitem{rb} T.~Vachaspati, arXiv:0802.0711 [astro-ph].

\bibitem{cps} R.~Khatri and B.~D.~Wandelt, Phys.\ Rev.\ Lett.\ {\bf
  100}, 091302 (2008).

\bibitem{magneto} D.~Battefeld, T.~Battefeld, D.~ H.~Wesley and
  M.~Wyman, ´ JCAP {\bf 0802}, 001 (2008).

\bibitem{cmb1} A.~A.~Fraisse, C.~Ringeval, D.~N.~Spergel and
  F.~R.~Bouchet, Phys.\ Rev.\ D {\bf 78}, 043535 (2008).

\bibitem{lensing1} K.~Kuijken, X.~Siemens and T.~Vachaspati,
  arXiv:0707.2971 [astro-ph].

\bibitem{lensing2} M.~A.~Gasparini, {\sl et. al.}, arXiv:0710.5544
  [astro-ph].

\bibitem{Davis:2008kg}
  A.~C.~Davis, W.~Nelson, S.~Rajamanoharan and M.~Sakellariadou,
  JCAP {\bf 0811}, 022 (2008)

\end{thebibliography}
\end{document}